\documentclass[12pt,onecolumn,showpacs,amsmath,amssymb,aps,nofootinbib,floatfix]{revtex4-2}
\usepackage{tikz}
\usetikzlibrary{calc,trees,positioning,arrows,chains,shapes.geometric,  decorations.pathreplacing,decorations.pathmorphing,shapes,  matrix,shapes.symbols}
\usepackage{epsfig}
\usepackage{xcolor}
\usepackage{amsmath}
\usepackage{amsmath}
\usepackage{amssymb}
\newcommand{\be}{\begin{eqnarray}}
\newcommand{\sqrtg}{\sqrt{-g}}
\newcommand{\ensemble}[1]{\left\{ #1 \right\}}
\newcommand{\probab}[1]{\mathcal{P}\ensemble{#1}}

 \newcommand{\secref}[1]{Section.\,(\ref{#1})}
\newcommand{\order}[1]{ \mathcal{O} \left( #1 \right) }
\newcommand{\lnz}{\ln \mathcal{Z}}

\newcommand{\ee}{\end{eqnarray}}

\newcommand{\ave}[1]{\left\langle #1 \right\rangle}
 \newcommand{\eqn}[1]{Eq.\,(\ref{#1})}
\newcommand{\eqcomma}{\phantom{AA},\phantom{AA}}

\begin{document}
\title{Gaussian generally covariant hydrodynamics}
\author{G.M.Sampaio}
\email[Corresponding author: ]{g224225@dac.unicamp.br}
\affiliation{Universidade Estadual de Campinas - Instituto de Fisica Gleb Wataghin\\
Rua Sérgio Buarque de Holanda, 777\\
 CEP 13083-859 - Campinas SP\\
}
\author{G.Rabelo-Soares}
\email[Corresponding author: ]{grsoares@ifi.unicamp.br
}
\affiliation{Universidade Estadual de Campinas - Instituto de Fisica Gleb Wataghin\\
Rua Sérgio Buarque de Holanda, 777\\
 CEP 13083-859 - Campinas SP\\
}
\author{G.Torrieri}
\email[Corresponding author: ]{torrieri@ifi.unicamp.br}
\affiliation{Universidade Estadual de Campinas - Instituto de Fisica Gleb Wataghin\\
Rua Sérgio Buarque de Holanda, 777\\
 CEP 13083-859 - Campinas SP\\
}
\begin{abstract}
We develop a version of fluctuating relativistic hydrodynamics in a way very different from the usual derivation: Instead of treating it as a coarse-grained deterministic theory expanded in gradients of equilibrium quantities, we treat it as a stochastic theory, characterized by partition functions in each cells, expanded in cumulants.   We show that the Gaussian ansatz allows us, via the gravitational Ward identities acting as a constraint between the variance and the average, to maintain full general covariance, with hydrodynamic flow emerging as an approximate Killing vector.   If the symmetry of ideal hydrodynamics, volume-preserving diffeomorphisms, is preserved, we show that linear response formulae are also generally covariant.   We discuss our results and argue that in this approach, the applicability of the effective theory is parametrized around a very different quantity than the Knudsen number, offering hope of understanding the applicability of hydrodynamics to small systems.
\end{abstract}
\maketitle
\section{Introduction \label{secintro}}
The observation of hydrodynamics in heavy-ion collisions (for a review, see \cite{kodama,jeon2}) has rekindled interest in the study of relativistic hydrodynamics as an effective theory and the derivation of it from microscopic theories
\cite{disconzi,kuboreview,teaney}.   These approaches are very sophisticated, but at their base is the idea of describing dynamics by coarse-grained averages of conserved quantities evaluated close to thermal equilibrium.  Neglecting the chemical potential (the only conserved quantity is the energy density), this means that 
\begin{equation}
\label{usualhydro}
\partial^\mu \ave{T_{\mu \nu}}=0 \eqcomma \ave{T^{\mu \nu}}= T_0^{\mu \nu}+\Pi^{\mu \nu} \eqcomma \partial_\mu \left(s u^\mu\right)\geq 0 \eqcomma Ts \sim \Pi_{\mu \nu} \partial^\mu \beta^\nu.
\end{equation}
The first relations are conservation equations driving the dynamics and the last represents the constraint of the second law of thermodynamics.

One then assumes the main part of the energy momentum tensor $T^{\mu \nu}_0$ in some frame defined by the flow $u_\mu \equiv T\beta_\mu$  is described by an equilibrium expectation value
\begin{equation}
\label{equil}
T_0^{\mu \nu}=\Lambda_\mu^\alpha \Lambda_{\nu}^\beta \left(\mathrm{diag}\left[e,p(e),p(e),p(e)\right]\right)_{\alpha \beta}\eqcomma p=-T\lnz\eqcomma e=\frac{\partial \lnz}{\partial \beta},
\end{equation}
and the rest obeys a relaxational equation, defined by a time-scale $\tau_\pi$ whose fixed point is a gradient multiplied by a viscosity.  \textcolor{black}{In a gradient expansion to first order this is, schematically, of the form}
\begin{equation}
\mathcal{D}\left( \tau_\pi,T \right) \Pi_{\mu \nu} + \Pi_{\mu \nu}  + \order{\mathcal{D}^{2},\Pi^{2},T^{2},u^{2}}= \eta \times \order{\partial u,\partial T}+\order{\partial^{2}},
\end{equation}
(the dissipative dynamics leads to the second law of thermodynamics being generally satisfied, with entropy growing in the co-moving frame $\dot{s} \sim \Pi_{\mu \nu}\partial^\mu \beta^\nu$).

This series can be regarded as an effective field theory expansion whose coefficients such as $\eta,\tau_\pi$ can be computed from Kubo formulae.  The system then becomes a closed system of equations that can be integrated from the initial conditions for $T,u_\mu,\Pi_{\mu \nu}$.

However,there are some reasons to think that this approach is missing important aspects, both experimental and theoretical.
On the experimental side, the observation of hydrodynamics in systems of only a few degrees of freedom  \cite{nagle} poses a challenge to the theories (the Boltzmann equation and AdS/CFT) most often used as ultraviolet cutoffs to hydrodynamics:   The first is based on the so-called Grad limit (density goes to infinity and scattering rate to zero in such a way that their product remains finite) while the second is based on the `t Hooft limit (number of colors squared goes to infinity and coupling constant goes to zero in such a way that their product is large).  While these approaches are very different one has in both cases a hierarchy\footnote{Note that $s^{-1/3}$ reflects the distance between degrees of freedom while $\eta/(sT)$ refers to the sound wave dissipation length.  Thus, this hierarchy could to strongly correlated systems where a Boltzmann description is inapplicable. For example, it applies to water, whose molecules are localized by hydrogen bonds,but the intermolecular distance is parametrically smaller than the sound wave dissipation length. \textcolor{black}{It could also apply to a collisionless Vlasov equation where the system looks chaotic on the scale of the wavelength of a typical perturbation but fluctuations are negligible on this scale, as in \cite{functional}}} \cite{scales}
\begin{equation}
\label{scales}
s^{-1/3}\ll \frac{\eta}{sT} \ll \left(\partial u\right)^{-1},
\end{equation}
the first inequality, which both the Grad and the `t Hooft limit assume to be small, seems experimentally falsified by \cite{nagle}.

On the theoretical side, the connection between hydrodynamics and statistical mechanics is largely mysterious.    The ambiguity around the definition of flow $u_\mu$ away from perfect equilibrium \cite{disconzi} muddles the connection between hydrodynamics and statistical mechanics, defined around the maximum entropy state in a certain frame \cite{palermo} in statistical mechanics but amenable to field redefinitions if hydrodynamics is seen as a classical field theory \cite{kovlec}.   

One way to make sense of this is that corrections to Navier-Stokes \cite{geroch,pretorius,kovreg,spalinski} are ``regulators'' rather than physical effective theory terms: when ambiguities are ``small'', the effective theory does not matter, while when they are large, hydrodynamics should not be used as an effective theory at all.   The phenomenological success of hydrodynamics at large gradients and the sophistication of the effective theory expansion make such an interpretation arguable. 

A different understanding (going back to \cite{is}, see \cite{masoud} for a recent review) is that rather than requiring quasi-stationarity or a local entropy maximization, local equilibrium just requires a mapping between each volume cell and a global equilibrium state, through which the approach to equilibrium is determined.   While this might make it easier to rationalize the applicability of hydrodynamics for large deviations from equilibrium, it is difficult to relate this picture to a distribution of microstates.  

Thus, the prescriptions to calculate fluctuations \cite{llfluct} and higher cumulants seem to be related to fundamental statistical mechanics (defined by the maximization of entropy under conservation law constraints) only in certain limits (relatable to Eq. \ref{scales}) \cite{xinan,skokov,mullins1,mullins2,glorioso,eightfold,akash,singh}.   \textcolor{black}{In most such approaches the cumulant and gradient expansion are two expansion parameters, \cite{xinan}, which has obvious tensions not just with fluctuation dissipation but also with the definition of local equilibrium}.
To illustrate the latter ambiguity, while the global equilibrium state can be defined for arbitrary flow fields as an entropy maximum \cite{palermo}, we know nothing in general about its Gibbs stability structure and hence about the effect local fluctuations have on it`s further development (is the ``equilibrium'' stable?).   Consequently, 
$\Pi_{\mu \nu}$ and, more worryingly, the entropy content, are similarly ambiguous.

Moreover, it is well known that the derivations of the transport coefficients from thermal field theory are plagued by dubious convergence and divergences \cite{parnold,moore}.  As a related point, when backreactions of fluctuations are included using state-of-the-art field theory techniques \cite{glorioso} results in a loss of universality of hydrodynamics \cite{akash} through the appearance of ``long-lived hydrodynamic modes'' and ``stochastic transport coefficients'' \cite{akash,gavlong,danhonilong}.  This is in direct contrast with the fact that close-to-ideal hydrodynamics seems to appear everywhere we look in experiments and begs the question of whether such coefficients and modes are physically detectable.

In this work, we shall continue an earlier research program \cite{grozdanov,montenegro,ghosts,crooks,functional,ergodic,gauge,randompol} that seeks to clarify these issues by re-deriving hydrodynamics from a different approach, centered on imposing the deep symmetries of hydrodynamics and connecting them to statistical mechanics in a way that fluctuations are also included from the start along with dissipation.   This means the object that is usually thought of as being the main driver of the evolution, the flow vector $u_\mu$, plays no explicit role in the dynamics, consistently with it's lack of unique definition away from perfect equilibrium. Furthermore,
$\Pi_{\mu\nu}$ does not appear either, but rather is replaced by a higher-rank object describing fluctuations and propagation.

It is widely known that {\em ideal} hydrodynamics has deep similarities to general relativity \cite{cov1,cov2,cov3}, being generally covariant w.r.t. volume-preserving diffeomorphisms (the largest group of possible diffeomorphisms not involving time \cite{son}), or equivalently the ergodic limit under foliations $\Sigma_\mu$ \cite{ergodic}.
\textcolor{black}{Physically, this general covariance is reasonable: statistical mechanics is based around a quantity (grand canonical and microcanonical entropy $\lnz$, which we shall explicitly define in \eqn{zubpart} and \eqref{uideal}, respectively) which is related to the number of accessible microstates to the system.  Since the constraints on this accessibility are conserved currents ($T_{\mu \nu}$ and $J_\mu$), it is reasonable that $\lnz$, or at least $\frac{\delta \lnz}{\delta \ensemble{T_{\mu \nu},J_\mu}}$ be covariant under local transformations (the start of an isentropic expansion changes the system from an inertial to a non-inertial frame but should not change the total number of accessible states), where ``local'' is defined on the scale $l_{\text{micro}}$ of \eqn{scales}.  In the ideal limit, the physical meaning of the volume-preserving diffeomorphism invariance is exactly to allow a definition of a conserved current, the entropy, represented by the volume of the cells}.   Away from the ideal limit, the system is expected to be generally covariant for foliations ``smooth'' w.r.t. the scale at which the system becomes thermalized (in the eigenstate thermalization hypothesis, the density matrix becomes close to random) \cite{ergodic}.  

What was always unclear \cite{llfluct} is how to implement this general covariance away from the ideal limit.    It is obvious that Schwinger-Keldysh type theories \cite{glorioso} or theories with gradient expansion \cite{kuboreview} can {\em never} be generally covariant \cite{singh} because they are implicitly based on the assumption that the space gradient is parametrically smaller than the time gradient (Kubo formulae and Schwinger Keldysh expand around the asymptotic limit in space to determine the evolution in the next time-step).   
This expansion is inherently non-covariant and also fails if the fluctuation time-scale is larger than the mean free path.  In other words, the foliations are smooth not on the thermalization scale but rather on the gradient scale.  \textcolor{black}{when cumulant and gradient expansions inter-twine \cite{xinan} this problem is aggravated, since fluctuations around the local thermostatic state generate sound waves that interact on the gradient scale order-by-order.  From the phenomenological point of view, such an expansion is ill-posed: since the background is not experimentally measurable,  the two parameters cannot be uniquely determined from the data. }

We therefore try to expand not in gradients but in statistical cumulants, assuming local near-equilibrium.     At first sight, a linearized expansion in gradients and a cumulant expansion should coincide, but this is not the case. In the former, one assumes {\em the background} to be deterministic and {\em perturbations} to be a stochastically evolving ensemble.   Thus, the initial data are ``a function'' (initial energy density and charge distribution), the propagator is of closed form, calculable from a thermal background, and independent of the average.  It is clear that this has no chance of being generally covariant.    The latter approach treats both background and perturbations as evolving ensembles (the ``background'' ensemble is parametrized by the Gaussian partition function).   
As can be seen, for example in \cite{kovner,renoprob} such a Gaussian approximation can be a powerful tool even for strongly coupled problems once the fundamental symmetries are known and incorporated into the variance.

In this respect, the general covariance can then be implemented at the level of the ensemble (different foliations are different ``cuts'' of it).  The ``price'' is that the initial conditions must be an ensemble, the ``initial propagator'' must be read from this ensemble and the dynamics is encoded in the relation between the propagator and the average given by the Gravitational Ward Identity (a $\delta-$function partition, i.e. deterministic hydro, would not obey the Ward identity at all times, in a more complicated partition function than a Gaussian the Ward identity would not determine the dynamics.  In this respect Gaussianity is a sweet spot, although it can be a good approximation even for strongly coupled systems \cite{kovner})

The idea is that once you assume Gaussianity of the partition function at all times, all dynamics has to be described by two parameters, the average and the variance, the latter related to the equal-time two-point function.
These two parameters are exactly what are linked by the gravitational Ward identity, which therefore {\em must be} the equation that determines the dynamics.  Finally, an approximate equilibrium with fluctuations predisposes to a fluctuation-dissipation relation like the linear response one \cite{tong,forster,kadanoff,bonanca}, which must also be compatible with general covariance.

In the next sections, we elaborate on each of the above statements in detail.  In \secref{seczub} we will show how general covariance can be obtained by combining the Zubarev partition function \cite{zubarev} with the gravitational Ward identity \cite{peskin,boulware,jeon2}.  In \secref{secgauss} we will show that the Gaussian approximation to the partition function retains these features while keeping the system soluble exactly.   The dynamics, based around local equilibrium linear response theory \cite{tong,forster,kadanoff,bonanca} or equivalently the Crooks fluctuation theorem \cite{crooks,crooks2}, will then be examined in \secref{seclin}, where it will be argued that under certain assumptions this dynamics is generally covariant despite the special role of time.  Finally in \secref{secfinal} we shall discuss the implications and extensions of these ideas, focusing on what they imply for the applicability of hydrodynamics as an effective theory w.r.t. statistical mechanics \cite{ll,huang,khinchin,berry,qchaos,grozanal,universality,framid,gavinfo,jaynes,biro,jaynesgibbs} and in small systems \cite{nagle}, as well as non-relativistic limits \cite{kaminski,arnold,spont1,spont2,fluidsym,incompress,gaugemech}as well as extensions to hydrodynamics with spin (in particular the pseudogauge issue)  \cite{spinreview,gursoy,jeonspin,brauner,florkfluct,montecausal} and curved spacetimes \cite{visser,feyn,unimod,jacobson,basso,crookscov,megrav1,megrav2}.
\section{Symmetries of ideal hydrodynamics and covariant statistical mechanics \label{seczub}}
To deviate from equilibrium, we need dissipation correlated with fluctuation of the energy-momentum conserved current.  To do this, we need to transition from the microcanonical ensemble used in \cite{ergodic} to an ensemble that allows for energy-momentum fluctuations.

We therefore construct a statistical ensemble around local equilibrium by using, at each time-step, the Zubarev partition function\footnote{For the rest of this work we put $J_\mu=0$ for simplicity, the extension to finite chemical potential is straight-forward} \cite{zubarev} (though a microcanonical limit might in fact be possible, to be investigated later \cite{bonanca}) 
\begin{equation}
\label{zubpart}
\mathcal{Z}= \mathrm{Tr} \exp \left[ -\int_\Sigma d\Sigma_\mu \left(\beta_\nu \hat{T}^{\mu \nu} + \mu \hat{J}^\mu\right) \right],
\end{equation}
if $\Sigma_\mu$ is a space-like 4-vector
\begin{equation}
\Sigma_\mu=(t(x),\vec{x})=(t(\phi_I),\phi_{I=1,2,3}),
\end{equation}
(in its own rest frame $\Sigma_\mu=(0,x,y,z)$) its exterior derivative (in its own rest-frame $d\Sigma_\mu=(dV,0,0,0)$) is time-like.  
which also implies a general metric,
\begin{equation}
\label{foldef}
g_{\mu \nu}= \frac{\partial \Sigma_\mu}{\partial \Sigma^\nu}   \eqcomma d^{3} \Sigma_\mu = d\phi_I d\phi_J d\phi_K \epsilon_{\mu \nu \alpha \beta}\epsilon_{IJK}  
\frac{\partial \Sigma^\nu}{\partial \phi_I}
\frac{\partial \Sigma^\alpha}{\partial \phi_J}
\frac{\partial \Sigma^\beta}{\partial \phi_K},
\end{equation}
In the ideal limit $\phi_I$s are also invariant under volume-preserving diffeomorphisms 
\begin{equation}
\label{volpres}
\phi_I \rightarrow \phi_I'(\phi_J) \eqcomma \mathrm{det}_{IJ} \frac{\partial\phi_I}{\partial \phi_J'}=1,
\end{equation}
this symmetry group insures there is only a unique choice of the ``killing vector'' perpendicular to all $\phi_I$ in any diffeomorphism, which can be used to define the hydrodynamic flow $u_\mu$ and, in the ideal limit, its conserved entropy current $s u_\mu$ (normalized to a microscopic scale $T_0$)
\begin{equation}
\label{uideal}
u_\mu \partial^\mu \phi_{I=1,2,3}=0 \eqcomma u_\mu u^\mu = -1 \eqcomma  s u_\mu= T_0^{3}\epsilon_{IJK} \epsilon_{\mu \nu \alpha \beta} \partial^\nu \phi_I \partial^\alpha \phi_J \partial^\beta \phi_K,
    \end{equation}
but this symmetry is believed to be broken beyond ideal hydrodynamics \cite{grozdanov,montenegro}. Consistent with this supposition, there is no unique definition for $u_\mu$, with proposals ranging between the Landau frame (where $u_\mu$ is parallel to the density of entropy), the Eckart frame (where $u_\mu$ is parallel to the conserved charge), choices motivated by first-order causality (BDNK), and so on (for a review see \cite{disconzi}).   The effective theory expansion for these frames is slightly different, although these effects are thought to be subleading \cite{geroch}.
The conclusions of \cite{geroch}, numerically confirmed in works such as \cite{pretorius}, are that the causal completion of the Navier-Stokes equations does not matter if the deviations from equilibrium are small, and when they become different, it signals that \textcolor{black}{one should perhaps use the UV completion, or at least correct the hydrodynamics with e microscopic theory, with a convergence rate strongly depending on the nature of the latter \cite{wagner}}.
\begin{figure}[h]
\epsfig{file=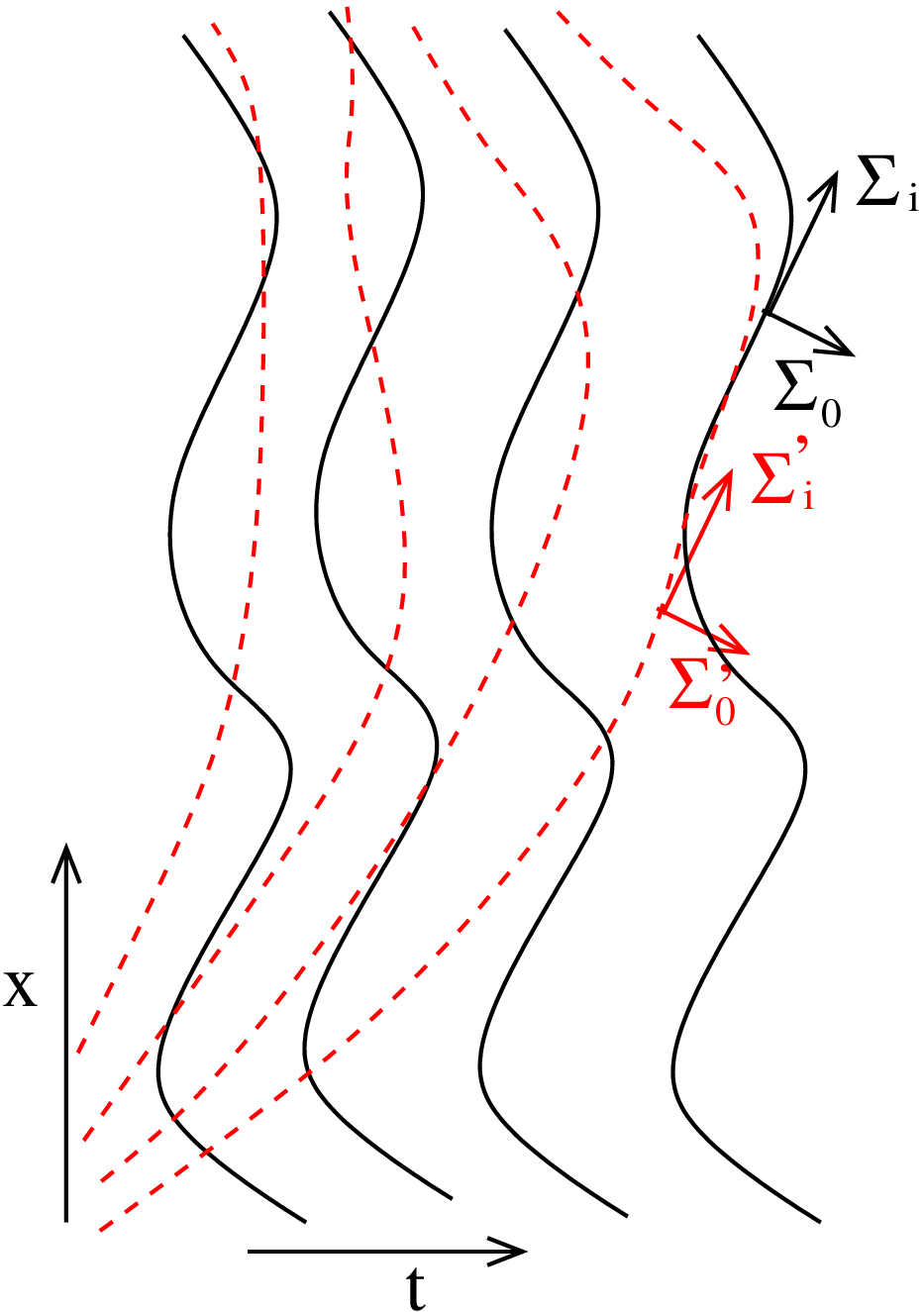,width=8cm}
\caption{\color{black}{Two foliations $\Sigma_\mu$ (shown in solid and dashed lines) representing two general coordinate systems.  Each metric is $g_{\mu \nu}=\partial 
\Sigma_\mu/\partial \Sigma^{\nu}$.
\label{figfol}}}
\end{figure}

Comparing the Zubarev approach with the above definitions, it is clear that different choices of $u_\mu$ correspond to different foliations $d\Sigma_\mu$ (see Fig. \ref{figfol}).
As can be seen in \cite{zubarev}, choosing a frame at rest with $t(x)$ and expanding gives you Kubo formulae.  However, \cite{zubarev} does not propose a dynamics different from the gradient expansion on which the above-listed approaches are based, while in \cite{zubarev} it is self-evident that $\Sigma_\mu$ is not a physical quantity and dynamics should not be based on it.
Complicating the situation is that while all foliations locally agree on causality, the same is not true at the nonlocal level (Fig. \ref{figfol}).  Space-like separated points under different foliations, whose local dynamics will eventually come into causal contact in the future, disagree on the temporal order of each other`s microscopic events (this is sometimes known as the ``andromeda paradox'' \cite{andromeda}).  Dynamics therefore should relate equal-time correlators (local fluctuations) to causal dissipative evolution in such a way that general covariance is preserved.

Given that, as shown in \eqn{foldef} $\Sigma_\mu$ defines a metric $g_{\mu \nu}=\partial \Sigma_\mu/\partial \Sigma^\nu$, it is promising that imposing general covariance (in the same sense of general relativity, although not involving ``the second invariant'', curvature, but just the first one, $\sqrt{-g}$, will lead to the right dynamics.

For deterministic ideal hydrodynamics, $\phi_I$ is defined up to a set of diffeomorphisms \cite{cov2,cov3}.  If hydrodynamics is non-ideal and fluctuating, you will have an ensemble of choices of $\phi_I$ that includes all these diffeomorphisms, as well as all possible choices of the flow vector $u_\mu$.  As discussed in \cite{gauge}, this can be thought of as a non-Abelian symmetry, so the separation of the two generally becomes non-trivial.  It becomes fundamental to understand how the partition function \eqn{zubpart} transforms under different choices of $\phi_I,u_\mu$.

Therefore, one can try to use the general covariance \cite{gauge} (the ambiguity of the choice of $\Sigma_\mu$ to remove the ambiguity of $\beta_\mu$ away from the perfect equilibrium.    Close to perfect equilibrium,  fluctuation dissipation enforces this general covariance in a way already demonstrated for the ideal case microcanonical limit \cite{ergodic} (\eqn{zubpart} is the Gran-Canonical limit of this).
In the limit of many degrees of freedom per unit volume of one mean free-path side, this Gaussian approximation should coincide with the usual Kubo/Schwinger/Keldysh relations.
We do, however, continue to describe the system as a partition function defined in each cell (which implicitly assumes that microstate number is localized and extensive, i.e. correlations are local and quantum correlations negligible, i.e. any Bell inequality violation in the microscopic degrees of freedom is negligible).

This means that at any given time. where the energy momentum tensor is generated by a metric perturbation, $\forall x,t \exists \mathcal{Z}(x,t)$ s.t.
\begin{equation}
  \ave{T^{\mu \nu}(x,t)}= \frac{\delta}{\left. \delta g_{\mu \nu}\right|_{x,t}} \lnz \eqcomma  \ave{\underbrace{T^{\mu \nu}(t_1,x_1)... T_{\mu \nu}(x_n,t_n)}_{n}}= \frac{\delta^n}{\delta \left. g_{\mu \nu} \right|_{x_1,t_1}... \delta \left. g_{\mu \nu} \right|_{x_n,t_n} } \lnz,
  \label{dpartfunc}
\end{equation}
\textcolor{black}{$n=2$ is what will be called $G_{\mu \nu \alpha \beta}(\Sigma_\mu \Sigma_\nu')$}.

To proceed and calculate the evolution at the partition function level, we use the relation between the partition function $\mathcal{Z}$ the Green function $D(x-y)$ and the field configurations $\phi(x)$ related by \cite{peskin}
\begin{equation}
\label{greendef}
\mathcal{Z} \propto \mathrm{Det}_{\phi} \left[ \phi(x) D(x-y) \phi(y) \right].
\end{equation}
For usual theories, $D(x-y)$ will be determined by the Lagrangian, but for a theory with both general covariance and local equilibrium, this will be impossible.   However, the Gravitational Ward identity, which automatically implements general covariance \cite{boulware,jeon2}, will give a constraint on the Green functions. It is\footnote{we note that, consistently with our definition of local equilibrium given in the previous section, we are not using doubled variables, which would require a more complicated version of the Ward identity. For thermal field theory, the Ward identity would be defined with Matsubara modes rather than Schwinger-Keldysh \cite{jeon2}.}
\begin{equation}
\label{ward}
\nabla_\mu\left\{ G^{\mu \nu \alpha \beta}\left( \Sigma_\mu,\Sigma'_\nu \right)- \frac{1}{\sqrtg}\delta\left(\Sigma'-\Sigma \right) \left( g^{\beta \mu}\left\langle\hat{T}^{\alpha \nu}\left(x^{\prime}\right)\right\rangle_{\boldsymbol{\Sigma}}+g^{\beta \nu}\left\langle\hat{T}^{\alpha \mu}\left(x^{\prime}\right)\right\rangle_{\boldsymbol{\Sigma}}-g^{\beta \alpha}\left\langle\hat{T}^{\mu \nu}\left(x^{\prime}\right)\right\rangle_{\boldsymbol{\Sigma}} \right)\right\}=0,
\end{equation}
where $\nabla_\mu$ is the covariant derivative ($\nabla_\mu g_{\mu \nu}=0$).

\textcolor{black}{The Ward identity can be derived from the requirement that the $n=2$ covariant derivative of \eqn{dpartfunc} vanishes \cite{jeon2}. Since $T_{\mu \nu}$ generates local metric perturbations, it can also be seen as the identity that reinforces the covariance of the propagator of $T_{\mu \nu}$ \cite{boulware}.  Higher order generalizations are possible but they will not be relevant at the Gaussian order. }
It is clear that for a non-fluctuating hydrodynamics $G_{\alpha,\beta,\gamma,\rho}$ vanishes and this equation reduces to the conservation of the energy-momentum current. 
\section{A generally covariant Gaussian ansatz \label{secgauss}}
We are now ready to examine the Gaussian expansion of \eqn{zubpart}.  In this limit, 
the form of $G_{\sigma \tau \mu \nu}$ will be
\begin{equation}
\label{gaussiang}
 G_{\sigma \tau \mu \nu}=\mathcal{P}_{\mu \nu, \alpha \beta}\left(\Sigma^\mu\right) f\left(\beta,T_{\mu \nu} T_{\alpha \beta}',\Sigma\right) ,
\end{equation}
where $f(...)$ in near-equilibrium has a Gaussian form, where $\Sigma(x)$ and $\Sigma(x')$ are two different points on the same foliation 
\begin{equation}
\label{gaussianansatz}
    f(...) \sim \prod_{\Sigma(x),\Sigma(x')} \exp \left[-\frac{1}{2}\left(T_{\mu \nu}(\Sigma(x'))-\ave{T_{\mu \nu}(\Sigma(x'))}\right)  C^{\mu \nu \alpha \beta}(\Sigma(x),\Sigma(x'))\left(T_{\alpha \beta}(\Sigma(x))-\ave{T_{\alpha \beta}(\Sigma(x))}\right)\right],
\end{equation}
dictated by a steepest descent prescription, that is, the first derivative $\left. \frac{\partial \lnz}{ \partial_\beta} \right|_0 \simeq 0$ because of local equilibrium, and all dynamics is encoded in the second derivative (this also leads to fluctuation-dissipation relations).   It will be 
\begin{equation}
\label{lnzexpansion}
 \lnz \simeq \left. \lnz \right|_0 - \textcolor{black}{\int_{\Sigma,\Sigma'}}\left. \frac{\partial^{2} \lnz}{\partial \beta_\mu \partial \beta_\nu} \right|_0 \left[ \ d\Sigma_\alpha  d\Sigma'_\tau \left( T^{\mu \alpha}(\Sigma)-\ave{T^{\mu \alpha}(\Sigma')} \right)\left( T^{\nu \tau}(\Sigma) - \ave{T^{\nu \tau}(\Sigma')} \right)\right].
\end{equation}
Provided the weak energy condition \cite{visser} is satisfied ($\forall d\Sigma_\mu,T_{\mu \nu}d\Sigma^\mu d\Sigma^\nu \geq 0$), any time-like $d\Sigma_\mu$ will lead to a negative exponent in \eqn{lnzexpansion} and hence to a well-defined Gaussian.  Note that for equation \eqn{lnzexpansion} to hold, $\beta_\mu$ must obey the Killing equation $\nabla_\nu \beta_\mu+\nabla_\mu \beta_\nu=0$ (or the first derivative would be non-zero).

The form of the projector $\mathcal{P}_{\alpha \beta,\mu \nu}$ is in the hydrostatic limit proportional to the width of the graviton propagator \cite{feyn}
\begin{equation}
\mathcal{P}_{\alpha \beta,\mu \nu} \propto 
C_{\alpha \beta \mu \nu}, \eqcomma \left. \mathcal{P}_{\mu \nu, \alpha\beta} \right|_{static}\rightarrow
g_{\mu \nu}g_{\alpha \beta}+g_{\mu \alpha} g_{\nu \beta}-g_{\mu \beta } g_{\nu \alpha}.
\end{equation}
\textcolor{black}{Roughly $C_{\mu \nu \alpha \beta}$ and $\ave{T_{\mu \nu}}$ correspond, in \eqn{greendef}, to $D(x-y)$ and $\ave{\phi}$.}

Performing a Gaussian functional integral of the form of \eqn{gaussianansatz} looks daunting, but one must remember that since $T_{\mu \nu}$ is symmetric and $C_{\alpha \beta \gamma \rho}$ is symmetric in all indices we can reduce it to a doubly contracted rank 4 tensor (let $\lambda_{i=1,4}$ be the eigenvalues of the first $T_{\mu \nu}$ and $\mu_{i=1,4}$ of the second one),
\begin{equation}
\label{cgaussian}
C_{\alpha \beta \gamma \rho } \left( T^{\alpha \beta} - \ave{T^{\alpha \beta}}\right)\left( T^{\gamma \rho}-\ave{ T^{\gamma \rho}}\right) \rightarrow C'_{\alpha \beta \gamma \rho } \times \text{diag} (\lambda_{1,2,3,4} ) \times \delta^{\alpha \beta}\times \text{diag} (\mu_{1,2,3,4} ) \times \delta^{\gamma \rho} ,
\end{equation}
remembering that $C_{\alpha \beta \gamma \rho }$ is a function defined in the position space, with a certain value at every point.

Furthermore, assuming that fluctuations around the average are `` small'', the boost representing them can be decomposed this way,
\begin{equation}
    T^{\mu\nu} -\ave{T^{\mu \nu}}= \Lambda^\mu_{\;\alpha}\, \text{diag}(\lambda_{1,2,3,4})^{\alpha\beta} \,\Lambda^{\;\nu}_{\beta} \eqcomma \Lambda^\mu_{\;\nu} \textcolor{black}{=\exp\left[  \int \frac{i}{2}  d\omega_{\alpha \beta} \left( M^{\alpha \beta} \right)^\mu_{\; \nu} \right] },
    \label{lorentzdec}
\end{equation}
where $M_{\alpha \beta}$ are the generators of the Lorentz group and $\Lambda$ diagonalizes the energy-momentum tensor (note that the metric $g_{\mu \nu}$ is the co-moving metric),
which reduces to a normal Gaussian integral.
In usual hydrodynamics, the first element of $\lambda$ corresponds to the energy density, the others to the pressure, and $\Lambda_{\mu \nu}$ changes the frame from the lab frame to `` co-moving frame''.
\begin{equation} \lambda_0 \simeq \mu_0 \simeq e \eqcomma \lambda_{1-3} \simeq \mu_{1-3} \simeq p \eqcomma \Lambda^\mu_\nu u^\nu \simeq (1,\vec{0}) ,
\label{diagframe}
\end{equation}
However, in general fluctuating hydrodynamics, $T_{\mu \nu}-\ave{T_{\mu \nu}}$ are elements of an ensemble characterized by fluctuations and local dissipation, so one expects $\lambda_i$ to be different (it obeys the universal eigenvalue distribution of a Gaussian random matrix ensemble; see discussion and references on random matrices \cite{functional}) and $\Lambda_{\mu}^{\nu}$ to have no physical interpretation in each ensemble.  Thus $C_{\mu \nu \alpha \beta}'$, the propagator in the reference frame where $T_{\mu \nu}-\ave{T_{\mu \nu}}$ has no simple analogue in the usual non-fluctuating hydrodynamics.

In fact, we have put ``co-moving'' in quotes because to use \eqn{lorentzdec} and \eqn{diagframe} to integrate \eqn{gaussianansatz} one has to make a very non-trivial assumption that has to be justified: That the fields $\Lambda_{\alpha \beta}$ in \eqn{lorentzdec} that diagonalize $T_{\mu \nu}-\ave{T_{\mu \nu}}$ are continuous and differentiable everywhere. 

If one interprets this $\Lambda_{\alpha \beta}$ as flow, it is clear that in ideal hydrodynamics with vorticity (at least with a continuus equation of state and no shocks) this {\em cannot occur}, which is another way of saying that for an ideal fluid with vorticity, in a flat space (Riemann tensor zero everywhere), $d\Sigma_\mu$ cannot be differentiable and parallel to $u_\mu$  everywhere.    

On the other hand, since, as recently noted in \cite{basso} (Appendix IV), the applicability of a covariant version \cite{crookscov} Crooks fluctuation theorem implies that in the absence of curvature entropy (i.e. $\lnz$) is generally covariant (this is also a consequence of the derivation in \cite{jacobson}), and since shocks generally change the entropy content \cite{ll}, the above assumption better hold.

So do we have a contradiction, and if so, how can we resolve it?  We note that the diagonalized $T_{\mu \nu}$ of an ideal non-fluctuating fluid is thrice degenerate $\mathrm{diag}(e,p,p,p)$ (the ambiguity of the three eigenvectors is, in fact, precisely the azimuthal part of the diffeomorphism group behind $\phi_I$ in \eqn{foldef} \cite{cov1,cov2,cov3}).   If one adds fluctuation and dissipation degenerate $T_{\mu \nu}$ form sets of measure zero, so we can safely assume the diagonalized $T_{\mu \nu}-\ave{T_{\mu \nu}}$ has the form 
$\mathrm{diag}(e,p_1,p_2,p_3)$, with three distinct uniquely defined eigenvalues that form a differentiable field in flat space.   This is a good indication that fluctuation-dissipation, when properly implemented, resolves the ambiguities inherent in ideal hydrodynamics with discontinuities.

The generalization of the Gaussian integral will therefore be (in the first equation the Einstein summation is disabled)
\begin{equation}
\label{gengauss}
\mathcal{Z} \simeq \int \prod_{\mu \nu} d\omega_{\mu \nu}  \exp \left[ - \omega_{\mu \nu} \omega^{\mu \nu}  C_{\mu \nu}^{ '\mu \nu}\right] \eqcomma C_{\mu \alpha}^{ ' \lambda \zeta}\delta_\alpha^\mu \delta_\beta^\nu=  \left( M^{\alpha \zeta} \right)^{\mu \nu} \left( M^{\beta \lambda} \right)^{\mu \nu} C_{\mu \nu \alpha \beta} ,
\end{equation}
\textcolor{black}{note that here we have expanded the Lorentz transformation to first order in phases. This is not necessary to find a representation of the form \eqn{cgaussian} (with diagonal energy momentum tensors only contracting the correlator), but terms beyond this order, $\order{\omega^2}$ and higher, will take the partition function away from its Gaussian form.
These should, however, be thought of as non-Gaussian corrections to the integration measure rather than the propagator itself (these corrections are logarithmic at most as ``large'' Lorentz transformations are suppressed by $\exp(\exp(-\omega^2))$).  
\textcolor{black}{For our model to be a fixed point in a lattice-type coarse-graining limit such a correction can only change the effective width, rather than introduce higher cumulants \cite{renoprob}}.
Just as a Gaussian ansatz with an appropriately tuned width seems to capture reasonably well deeply non-Gaussian lattice calculations \cite{kovner}, so we have confidence it will be qualitatively consistent here even when fluctuations are not negligible w.r.t. the average.
}

The following trick helps to proceed:  Given $\sqrt{J}_\mu$ and the tensor $(e_\mu)^\nu$ where $(e_\mu)^{0}=(1,0,0,0)$, $(e_\mu)^{1}=(0,1,0,0)$, etc., one can construct a diagonal matrix whose elements are given by $J_\mu$ (the squares of each element $\sqrt{J}_\mu$) as
\begin{equation}
\label{trick}
\left( \mathrm{Diag}(J_\mu) \right)_{\alpha \beta}=\sqrt{J}_\mu \sqrt{J}_\nu \left( e_\alpha \right)^\mu   \left( e_\beta \right)^\nu,
\end{equation}
($J_\mu \left( e_\alpha \right)^\mu$ is more compact but error-susceptible as one foregoes the Einstein summation convention).  The $e_\alpha$ is roughly corresponding to the Vierbein of the change in frame between the lab and the diagonalizing frame.

With this trick, the partition function at the point $\Sigma$ is
\begin{equation}
\label{gengausssolved}
\lnz(\Sigma)=\int_\Sigma d\lnz(\Sigma) \eqcomma \textcolor{black}{d \mathcal{Z} (d\Sigma)}=  \left( \sqrt{ \prod_i \lambda_i \mu_i}\sqrt{ C_{\alpha \beta}^{'\alpha \beta} }\right)^{-1} ,
\end{equation}
a note of caution is in order: $\lnz$ represents the density matrix, at each point in time according to a given choice of foliation $\Sigma_0$.   The RHS appears Lorentz covariant due to \eqn{trick}, but it contains two determinants (the eigenvalue products).  As a determinant is an inverse volume it should transform as a ``time'', matching the $\Sigma_0$ dependence of $\lnz$.   However, if the Crooks theorem is to be valid, $\lnz$ better be the same for two observers with different coordinate systems who coincide at $\Sigma_0$ and $\Sigma_0+d\Sigma_0$ \cite{basso,jacobson}, a requirement that the fluctuation-dissipation theorem motivated by linear response should satisfy.

With everything clarified, we can proceed.
Then one can use the Ward identity \eqn{ward} to obtain an equation of motion for $C^{\mu \nu \alpha \beta}$ in terms of $\Sigma$ (or equivalently, the metric).    The best way to do this is to integrate both sides of \eqn{ward} to eliminate the $\delta-$function. After some manipulation, using the famous identity 
\[\   \nabla_\mu A^{\mu} \equiv \frac{1}{\sqrtg} \partial_\mu \left( \sqrtg A^{\mu} \right), \]
it becomes \textcolor{black}{(the integral over $\Sigma'$ in \eqn{lnzexpansion} is absorbed by the $\delta-$function of \eqn{ward}) }
\begin{equation}
Q^{\alpha \beta}= \int d\Sigma_\mu  \nabla_\nu C^{\mu \nu \alpha \beta} \left[ 1+\frac{1}{\sqrtg}\sum_i \frac{1}{\lambda_i \mu_i\sqrt{C_{\zeta\delta}^{'\zeta \delta}}} \right],
\label{wardq}
\end{equation}
$Q_{\mu \nu}$ is a constant of the motion given by the initial conditions, although it is {\em not} an event-by event physical observable but rather one characterizing the whole ensemble.

Once this equation of motion is solved, one can then plug into \eqn{lnzexpansion} to find $\beta_\mu$ in terms of $\Sigma$. This would be a covariantization of Eq. 7.14 of \cite{huang}
\begin{equation}
\label{covarianthuang}
\ave{E^{2}}-\ave{E}^{2}\equiv C_V T^2  \Rightarrow \ave{\Delta T^{\mu \nu} \Delta T_{\alpha \beta}} \sim \left. \frac{\partial^{2} \lnz}{\partial \beta_\mu \partial \beta_\nu} \right|_0   d\Sigma_\alpha d\Sigma_\beta,
\end{equation}
\textcolor{black}{the above formula is closely related to the second derivative of the Helmholtz free energy, $-d^{2}\lnz/(dTdV)$, leading to the Maxwell relation $dS/dV=dP/dT$.  The covariantized version is}
\begin{equation}
\label{covariantmax}
\frac{\partial C_{\zeta \rho \mu \nu}}{\partial C`_{\zeta \rho \mu \nu}}\left( C_{\alpha \beta}^{'\alpha \beta} \right)^{-3/2} = \frac{\partial \beta_\mu \partial \beta_\nu}{\partial \Sigma_\mu \partial \Sigma_\nu},
\end{equation}
therefore, can give the $\beta_\mu$ field corresponding to the choice of foliation $\Sigma_\mu$.   In this sense, consistently with the discussion around \eqn{diagframe}, $C_{\alpha \beta}^{'\alpha \beta}$ has the information corresponding to the equation of state, while the ``off diagonal elements" of this complicated rank-4 tensor have the information corresponding to transport coefficients and fluctuations according to the Kubo formula \cite{teaney}. Thus, given the Fourier transform $\tilde{C}_{\alpha \beta \mu \nu}^{'}(k)$ 
\begin{equation}
\label{transcoeff}
\eta \sim k^{-1} \lim_{k \rightarrow 0} \text{Im} \tilde{C}_{xyxy}^{'}\phantom{A} c_s^2  \sim k^{-1} \lim_{k \rightarrow 0} \text{Re} \tilde{C}_{tttt}^{'} \phantom{A} \tau_\pi \sim k^{-2} \lim_{k \rightarrow 0} \text{Im} \frac{d}{dk} \tilde{C}_{xyxy}^{'}..., 
\end{equation}
with the detailed matching is left to subsequent work.  But we should underline that the full $\tilde{C}_{\alpha \beta \mu \nu}^{'}(k)$ is a field that evolves from the initial conditions of an ensemble of configurations of the $T_{\alpha \beta}$ field, not a functional of the Lagrangian alone.

Note that while in the hydrostatic case $C_V$ is purely a fluctuation, in the covariant case it will contain information of both local fluctuations and dissipative evolution, which needs to be constrained by a fluctuation-dissipation theorem.

\textcolor{black}{The philosophy we have been using in the derivation of this section has some similarity with the so-called ``density frame'' approach \cite{singh} in that we try to define the dynamics at the level of the partition function. In this respect, we believe that usually the frame in which $C'_{\alpha \beta \gamma \mu}$ is close to the density frame.   However, since, as clearly explained in \cite{singh}, in deriving the density frame one needs to explicitly give up covariance, we would not use this approach to calculate flow redundancies \cite{gauge}, which our covariant approach includes.   As argued in the discussion, these are essential for estimating the applicability of hydrodynamics.}
\section{General covariance of Linear response and Crooks fluctuation relations \label{seclin}}
The above discussion seems to be a non-starter as it appears to conflict with some basic facts of linear response theory \cite{forster,kadanoff,tong}.  As we know, fluctuations are governed by the equal-time behavior of the correlator, while dissipation is governed by the imaginary part.  Hence, while the two are related, this relationship is generally highly complex.  

The issue is that non-inertial refoliations, while they preserve causality locally, could lead to observers disagreeing on the causal order of space-like separated events, which will affect a given point in the future (Fig. \ref{figfol}). It is therefore difficult to see, for instance, how fluctuation-dissipation formulae derived in \cite{forster,kadanoff,tong} could have a $\mathrm{Re}\tilde{G}_{\mu \nu \alpha \beta}$ and $\mathrm{Im}\tilde{G}_{\mu \nu \alpha \beta}$ that transform in such a way that, for instance, $\tilde{G}_{\mu \nu \alpha \beta}(k,k')\tilde{T}^{\alpha \beta}(k) \tilde{T}^{\mu \nu}(k') $ is a scalar under general refoliation.   

On the other hand, \cite{forster,kadanoff,tong} assume that the background frame is precisely known and constant to the asymptotic scales. Precisely because of the ambiguity of causality at distant points in a foliation (Fig.\ref{figfol}), general covariance must emerge when this knowledge is absent.  In other words, local fluctuations of the background, when traced over the evolution of ensembles of configurations, must produce correlations at distant points that respect general covariance with respect to refoliations.

In this work, we shall assume that this is in fact the case and derive some consequences.   In the next section, we shall comment on the physical justification of this assumption.

 First we note that \cite{forster} uses the metric perturbation $\delta g_{\mu \nu}$ to define a ``source'' for an energy-momentum tensor.  To convert it into an initial condition, one can use the trick \textcolor{black}{ described in section 2.6 of \cite{forster} which we reproduce below in a ``covariant-like form'' ($t\rightarrow \Sigma_0$)}
\begin{equation}
\label{stepmetric}
\textcolor{black}{T_{\mu \nu}(\Sigma_0)= \int \exp\left[i\epsilon \Sigma_0'\right] G^{\mu \nu \alpha \beta}(\Sigma_0-\Sigma_0') \delta g_{\alpha \beta}(\Sigma_0')\,d\Sigma_0',}
\end{equation}
we start an infinitesimal source at an initial time
\begin{equation}
\delta g_{\mu \nu}=\left\{ 
\begin{array}{cc}
\delta g_{\mu \nu}e^{\epsilon \Sigma_0} & \Sigma_0>0\\
0 & \Sigma_0 \leq 0
\end{array}
\right. ,
\end{equation}
and use the covariantized Laplace transform.
Now, since we can use $G^{\mu \nu \alpha \beta}$ to construct a kernel to integrate one step in time
\begin{equation}
\label{step}
    \left\langle\hat{T}^{\alpha \beta}\right\rangle_{\Sigma+d \Sigma}=i \int \frac{1}{\sqrtg}\mathcal{G}^{\mu \nu \alpha \beta}(\Sigma'_0-\Sigma_0) \ave{T}_{\mu \nu} (\tau,x)\,d\Sigma_0 ,
\end{equation}
where $\mathcal{G}^{\alpha \beta \mu \nu}$ is the ``renormalized'' and time-ordered propagator 
\begin{equation}
\label{gtilde1}
\mathcal{G}^{\mu \nu \alpha \beta}(\Sigma_0,\Sigma_i)=\int \frac{1}{\sqrt{g_k}} d^{3}k_i \, \tilde{\mathcal{G}}^{\mu \nu \alpha \beta} e^{ i k_{i} \Sigma_i },
\end{equation}
\begin{equation}
\label{gtilde2}
 \tilde{\mathcal{G}}^{\mu \nu \alpha \beta}=\frac{1}{2i} \left( \frac{\tilde{G}^{ \alpha \beta \mu \nu}(\Sigma_0,k)}{\tilde{G}^{ \alpha \beta \mu \nu}(-i\epsilon\Sigma_0,k)}-1\right).
\end{equation}
Here we need to understand the covariance properties of \eqn{gtilde1}, \eqn{gtilde2}, \eqn{step} and \eqn{stepmetric}.
In each case, the LHS is covariant, so the RHS better be.  But it certainly does not appear this way, as it contains an integral in $\Sigma_0$.   Here we need to go back to the discussion of the covariance properties of \eqn{gengausssolved}.   Plugging the Gaussian ansatz into the definition of $\tilde{G}_{\alpha \beta \mu \alpha}$, we get
\begin{equation}
\label{gtildeab}
    \tilde{G}_T^{\mu \nu \alpha \beta}\left(\Sigma_0\right)=C^{\mu \nu \alpha \beta } \frac{2}{\lambda_i \mu_i C_{\alpha \beta}^{\prime \alpha \beta}} \int_{-\infty}^{\infty} e^{-A} \sin (B) e^{-i \omega^{\prime} \Sigma_0} \Theta\left(\Sigma_0\right) d \omega^{\prime},
\end{equation}
where
\begin{itemize}
    \item $A=\frac{\left(\Sigma_\mu k^\mu\right)^2}{2 \lambda_i \mu_i C_{\alpha \beta}^{\alpha \beta \prime}} $ real part, transforms as a volume via the determinant, thereby canceling the $\Sigma_0$ in the exponent;
\item $B=k \cdot\langle\Sigma\rangle $ imaginary part, transforms as a scalar.   
\end{itemize}
The second item is a Lorentz scalar, but the first has a denominator, which is not.   To understand this, we need to examine more carefully the relativistic 4-volume in general coordinates, 
\begin{equation}
\label{genvol}
    d V d t \rightarrow \sqrtg d^4 x \eqcomma \prod_i \lambda_i \mu_i C_{\mu \nu}^{'\mu \nu} \sim dV \times dt,
\end{equation}
$dV dt$ and $d^{4} x$ are obviously invariant under any local Lorentz transformation.   However, $\sqrtg$ is not, reflecting the non-invariance of the denominator of $A$ (curved $d\Sigma_\mu$ will generally introduce a non-trivial $\sqrtg$ which will change any factor of $dt \times dV$, reflecting the fact that $\lnz$ is a phase space volume, which in principle is not invariant under rescaling of coordinates.

However, as discussed in \cite{ergodic}, the symmetry defining ideal hydrodynamics are volume-preserving diffeomorphisms \eqn{volpres}, which is exactly the group that leaves $\sqrt{-g}$ invariant under rescaling of volume only (\cite{unimod}).  Local equilibrium means that the dissipative evolution is indistinguishable from an isentropic fluctuation at the local level (in the Gaussian ansatz, this is implemented by assuming local ergodicity at the level of the mean and the variance).   Hence, assuming volume-preserving diffeomorphisms in addition to local Lorentz invariance should ensure the general covariance of $A$, as required.  The covariance of \eqn{gtildeab} follows as contractions of $w$ and the denominator on the RHS follow. 

It is therefore clear how the fluctuation dissipation theorem \cite{tong}, relating the dissipative coefficient spectral function $\chi(w)$ and the fluctuation coefficient $S(w)$ by frequency mode $w$
\begin{equation}
\label{fluctdissform}
\chi(w) =-\frac{1}{2}(1-e^{-\beta w})S(w) ,
\end{equation}
becomes covariantized.  $\chi \sim \text{Im}\tilde{C}_{\alpha \beta \mu \nu}$ while $S(w) \sim \text{Re} \tilde{C}_{\alpha \beta \mu \nu}$.  Without volume-preserving diffeomorphisms, this relation could be Lorentz invariant ($\beta w \rightarrow p_\mu \beta^\mu$ where $p_\mu$ is the 4-momentum) but not generally covariant, due to the special role of $w$.
However, physically, volume-preserving diffeomorphisms leave invariant the entropy content of the cell and the proper time $\beta_\mu d\Sigma^\mu$ of its evolution.   Hence, as $p^\mu \beta_\mu \rightarrow T_{\mu \nu} \beta^\mu d\Sigma^\nu$, any volume-preserving rescaling of $w$ (the conjugate of proper time) will preserve \eqn{fluctdissform} term by term.   \textcolor{black}{ The physical significance of this is that away from the ideal limit, entropy is, of course, not conserved, but its violation over the scale of a coarse-graining volume element is indistinguishable from a fluctuation.   For this to be consistent, one needs to consider only those diffeomorphisms that leave the entropy invariant, which are exactly volume-preserving diffeomorphims \cite{ergodic}.  A $\partial \Sigma_\mu/\partial \Sigma_\nu$ change in the volume of a cell would also change its microscopic entropy, and hence we do not expect the fluctuation-dissipation theorem to be maintained for transformations of this type. }

\textcolor{black}{This realization also clarifies the role of the equation of state in our approach. It is simply the Lagrangian definition of the equation of state used in ideal hydrodynamics. 
 In terms of an arbitrary EoS $F(x)$, it is \cite{cov3}
\begin{equation}
\label{lageos}
e=-F(s) \eqcomma p=s^{2} \frac{de}{ds^{2}}\equiv -\frac{\lnz}{\sqrt{\beta_\mu \beta^\mu}}, 
\end{equation}
given an entropy and a metric $g_{\mu \nu}$ one can calculate the corresponding $\ave{T_{\mu \nu}}$ from \eqn{dpartfunc}, $u_\mu$ from \eqn{uideal} and $C_{\alpha \beta \gamma \mu},C_{\alpha \beta \gamma \mu}'$ from \eqn{covariantmax}, thereby closing the Ward identity equation in this coordinate system.}

Note that the more general effective theory expansion based on increase in entropy and gradients generally posits that volume-preserving diffeomorphisms are a property of ideal fluids only (see Appendix B of \cite{eightfold}).  Imposing general covariance and allowing fluctuations in entropy allow us to see that, as intuited from ergodic theory \cite{ergodic}, the role of these symmetries extend beyond this limit.    Physically, once fluctuations are taken into account in a generally covariant way, while $u_\mu$ ceases to be a Killing vector in the sense of \eqn{volpres}, this deviation becomes locally indistinguishable from a fluctuation, and this is encoded in the structure of the Gaussian width $C_{\mu \nu \alpha \beta}$.   This means that the ergodic hypothesis is still locally valid, \textcolor{black}{only covariance under volume-preserving diffeomorphisms is physically relevant in correlating $C_{\mu \nu \alpha \beta}$ to $\ave{T_{\mu \nu}}$. 
 But if one coarse-grains by tracking the average only this symmetry remains irrelevant}.

  The closed system of equations needed to evolve the hydro is shown in Fig. \ref{linearflux}.
 \begin{figure}
 \tikzstyle{block} = [rectangle, draw, text width=15cm, text centered, rounded      corners, minimum height=2cm]
 \begin{tikzpicture}
 [node distance=2.5cm,
 start chain=going below,]
\node (n1) at (0,0) [block]  {Start with a lattice with an ensemble $\ensemble{T_{\mu \nu}(x_i,t)}$ at each point of it};
\node (n2) [block, below of=n1] {Construct $\ave{T_{\mu \nu}(x_i,t)}$ and $C_{\mu \nu \alpha \beta}(x_i,x_j,t,t)$  and the corresponding $C_{\mu \nu \alpha \beta}'$ from it (\eqn{dpartfunc} and \eqn{covariantmax})};
\node (n3) [block, below of=n2] {Use General covariance and $C_{\mu \nu \alpha \beta}(x_i,x_{i+1},t,t)$ to construct $\mathcal{G}_{\alpha \beta \mu \nu}(x_i,x_{i+1},t,t)$ (\eqn{gtilde1}) };
\node (n4) [block, below of=n3] {Use linear response on each $\ensemble{T_{\mu \nu}(x_i,t)}$ to get  $\ensemble{T_{\mu \nu}(x_i,t+\Delta t)}$ \eqn{step}};
\node (n5) [block, below of=n4] {Calculate $\ave{T_{\mu \nu}(x_i,t+\Delta t)}$. Use the Ward identity to get $C_{\mu \nu \alpha \beta}(x_i,x_{i+1},t+\Delta t,t+\Delta t)$}; (\eqn{wardq}) 
\draw [->] (n1) -- (n2);
\draw [->] (n2) -- (n3);
\draw [->] (n3) -- (n4);
\draw [->] (n4) -- (n5);
\draw [->] (n5.east) -| ++(1,0) |- (n2.east);
\end{tikzpicture}
  \caption{The dynamics, defined via a numerical lattice, of the fluctuating hydrodynamic evolution using linear response. \label{linearflux}}
  \end{figure}
It is also possible that the techniques developed in \cite{singh} will allow us to bypass the linear response integral and the use of Kramers-Kronig analyticity.
Starting from \textcolor{black}{the Gibbs-Duhem relation} and remembering that both $\lnz$ and $\mathcal{P}$ should be Lorentz scalars, and using the definition of the latter from the speed of sound, we can obtain a covariantized version of the Gibbs-Duhem relation (Note the use of discrete derivatives $\Delta$ as we shall use stochastic calculus to set-up the dynamics)
\begin{equation}
\label{gibbsduhem}
-\Delta \lnz=- \beta_\nu J^\nu \Delta\mu+P^i \Delta\beta_i - \Delta\Sigma^0 \beta_0 \int_0^{P^{0}}c_{s}^{2}(e) de  \eqcomma P_{\alpha=0,i=1...3}\equiv T_{\alpha \beta} d\Sigma^\beta ,
\end{equation}
as discussed below \eqn{genvol} in analogy to \cite{ergodic} this expression can be made covariant under arbitrary foliations by imposing local ergodicity.  Flow then emerges as a Killing vector as a discretized version of \eqn{uideal} and \eqn{volpres}.
\begin{figure}
 \tikzstyle{block} = [rectangle, draw, text width=15cm, text centered, rounded      corners, minimum height=2cm]
 \begin{tikzpicture}
 [node distance=2.5cm,
 start chain=going below,]
\node (n1) at (0,0) [block]  {Start with a lattice with an ensemble $\ensemble{T_{\mu \nu}(x_i,t)}$ at each point of it};
\node (n2) [block, below of=n1] {Construct $\ave{T_{\mu \nu}(x_i,t)}$ and $C_{\mu \nu \alpha \beta}(x_i,x_j,t,t)$ from it};
\node (n3) [block, below of=n2] {Use \eqn{lageos} to find construct the entropy, and $C_{\mu \nu \alpha \beta}(x_i,x_{i+1},t+\Delta t,t+\Delta t')$ for a time-like foliation  };
\node (n4) [block, below of=n3] {Use the Crooks fluctuation theorem and acceptance/rejection to get  $\ensemble{T_{\mu \nu}(x_i,t+\Delta t)}$};
\node (n5) [block, below of=n4] {Calculate $\ave{T_{\mu \nu}(x_i,t+\Delta t)}$. Use the Ward identity \eqn{wardq} to get $C_{\mu \nu \alpha \beta}(x_i,x_{i+1},t+\Delta t,t+\Delta t^\prime)$}; 
\draw [->] (n1) -- (n2);
\draw [->] (n2) -- (n3);
\draw [->] (n3) -- (n4);
\draw [->] (n4) -- (n5);
\draw [->] (n5.east) -| ++(1,0) |- (n2.east);
\end{tikzpicture}
  \caption{The dynamics, defined via a numerical lattice, of the fluctuating hydrodynamic evolution using Crook's fluctuation theorem. \label{crooksflux}}
  \end{figure}
Hence, as long as $u_\mu$ is proportional to $d^{3} \Sigma_\mu$ up to a thermal fluctuation (in non-potential flows making it proportional everywhere will result in singularities), the resulting \eqn{gibbsduhem} will give a temperature close to the Gibbsian definition.

We shall now describe the dynamical evolution, based on covariant fluctuation-dissipation relations. According to Crook's theorem \cite{crooks,crooks2} one can define the dynamics stochastically, with the probability $\probab{P_\mu}$
\begin{equation}
\label{crooksdyn}
\frac{\probab{\left.P_\mu\right|_{\tau}\rightarrow \left.P_\mu\right|_{\tau+\Delta \tau }}}{\probab{\left.P_\mu\right|_{\tau+\Delta \tau}\rightarrow \left.P_\mu\right|_{\tau }}} \sim \exp[\left. \lnz \right|_{\tau+\Delta \tau}-\left. \lnz \right|_\tau] \eqcomma \Delta \tau=\beta_\mu \frac{\Delta^{3}\Sigma^\mu}{\Delta^{3}\phi_{i=1,2,3}},
    \end{equation}
    a numerical implementation of how this works is sketched in Fig. \ref{crooksflux}
    This is ``Gibbsian'' because, of course, event-by-event $d\lnz$ is not the ``real'' entropy, but a system evolved this way with a ``reasonable'' $\Sigma_\mu$ over many steps will achieve an $\lnz$ that is independent of the choice of $\Sigma_\mu$ and with an $\lnz$ approximating the ``real'' entropy.
    
    Then one can make a link with the Zubarev Gaussian ansatz found described in \eqn{gaussianansatz} if one has (in analogy to Eq. (11) of \cite{singh}) 
    \begin{equation}
    \left. \lnz \right|_{\tau+\Delta \tau}-\left. \lnz \right|_\tau \sim \exp \left[-\Delta_\mu \beta_\nu C^{\mu \nu \alpha \zeta} \Delta_\alpha \beta_\zeta \right] \eqcomma \Delta_\mu O \equiv \frac{\Delta O(x^\mu)}{\Delta x^{\mu}},
    \end{equation}
    with the same kernel  $C^{\mu \nu \alpha \zeta}$ used to define the $G_{\mu \nu \alpha \beta}$ of \eqn{gaussianansatz} (which, in turn, obeys the gravitational Ward identity \eqn{ward}).   

The relationship between the dynamics of the type \eqref{crooksdyn} and the linear response approach based on \eqref{step} mirrors the relationship between the Fokker-Planck and the Langevin equation
\begin{equation}
\label{langfok}
\frac{dx}{dt}\sim \mathrm{Im}\mathcal{G} (x-\ave{x})+\mathrm{Re}\mathcal{G} \hat{\delta} \Leftrightarrow F(x,t+\Delta t)= \underbrace{\int d x' \, \overbrace{ P(x,t+\Delta t|x',t)}^{\text{Gauss} \left(\ave{x},\sigma² \sim \mathrm{Re}\mathcal{G}+i \mathrm{Im} \mathcal{G} \right)} F(x',t)}_{\sim -\mathrm{Im}G \frac{\partial}{\partial x}F(x)+ \mathrm{Re}G \frac{\partial^{2}}{\partial x^{2}} F(x)},
\end{equation}
$P$ on the right-hand ``Fokker-Planck'' equation is the right hand side of \eqref{crooksdyn} while the drift and fluctuation Langevin terms are the real and imaginary part of the time-ordered commutator $\mathcal{G}^{\alpha \beta \mu \nu}$ of \eqref{step}.  In the former case, one has an equation of motion for $\lnz$, the `` ensemble'' of hydrodynamic configurations.  In the former case, it is an equation of motion for the first two cumulants of the operator corresponding to $\hat{T}_{\mu \nu}$.  As long as one maintains the Gaussian approximation, the two approaches should give the same result, although a formal demonstration of this conjecture is beyond the scope of this work.
    This picture also gives an intuitive way to understand why hydrodynamics might work better in strongly coupled small systems where the eigenstate thermalization hypothesis \cite{berry} holds but where fluctuations are large. The coarse graining in $\Delta \tau$ is on the scale of thermalization and isotropization, which according to the eigenstate thermalization hypothesis/Berry conjecture\cite{berry} could indeed be really short w.r.t. fluctuation timescale, so the number of DoF in each $\Delta_\mu \beta^\mu$ cell is $\order{1}$.
    In such a picture, the RHS of \eqn{crooksdyn} could be very close to unity, i.e. at each step $\Delta \tau$ entropy is nearly as likely to decrease as to increase.   Hence, over many $\Delta \tau$ the evolution is nearly isentropic, which would correspond to close to ideal hydro.   This can be verified by numerical simulations.  We should also note that recent considerations of a similar nature were made in the context of so-called maximally chaotic systems \cite{qchaos}.
    \section{Discussion \label{secfinal}}
\begin{figure}[h]
\epsfig{file=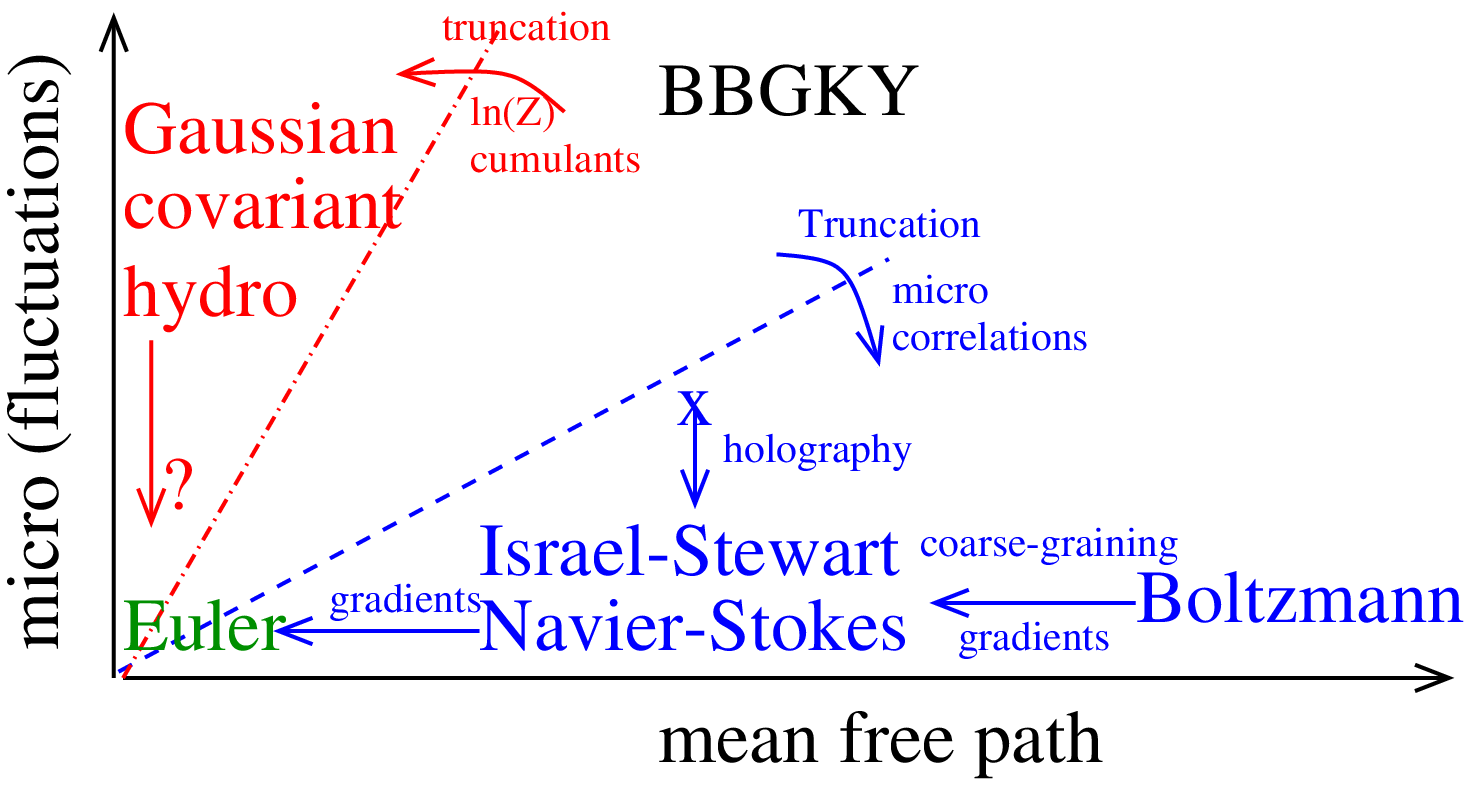,width=16cm}
\caption{Validity of different approaches to hydrodynamics from microscopic theory \label{figscales}.  ``BBGKY" refers to a general many body state with complicated correlations, see discussion in \cite{functional}.}
\end{figure}

    In this paper, we present a hydrodynamic limit that we argue is valid in a different region of the space of scales defined in \eqn{scales}.   This is shown in Fig. \ref{figscales}:  The approaches most commonly used in the literature, transport and holography \cite{kuboreview} arise from generic many-body theory\footnote{denoted by BBGKY because of the hierarchy, see 
\cite{functional} and references therein.}  when thermalization time and mean free path are sizeable but microscopic fluctuations are negligible.  In this limit general covariance is irrelevant since it occurs at the scale where it is inaccessible by macroscopic measurements.
    
    This allows to separate the background from a perturbation, assign a definite frame to the former and relate the latter to transport coefficients.  Our approach is argued to work in the opposite limit, where the thermalization time is negligible (perhaps because of eigenstate thermalization \cite{berry}) but fluctuations are large and therefore there are many ways to define a background, generally by a non-inertial foliation.  While within the approaches in \cite{kuboreview} we have a good understanding of how the near-ideal limit emerges from the Boltzmann and holographic limit in the lower panel delimited by the dashed line, the region of applicability of the Gaussian-Covariant effective theory (delimited by the dot-dashed line) and its approach to the ideal limit could yield surprises:  eigenstate thermalization could lead to a much larger angle than expected of the dot-dashed line in Fig. \ref{figscales} w.r.t. to the y axis than what a naive fluctuation-dissipation estimate ($\sim N_{dof}^{-1/2}$) would suggest.    
    
    Provided equilibration happens rapidly, the evolving ensemble could be centered around a low-viscosity evolution, in other words, good fluidity could arise in few-particle systems.  Arguments made previously \cite{gauge,functional} make this conclusion reasonable.   
    
Although the approach proposed in this work looks unusual, $g_{\mu \nu}$ is the conjugate variable to $T_{\mu \nu}$, it admits an acoustic frame ($\beta_\mu$ is its Killing vector), so there is no reason why hydrodynamics cannot be defined in terms of it.
What makes it unusual is that, since usually one implies a hydrostatic background (which we want to avoid) and uses gradients as a source of perturbation, what is above just a statement about the equation of motion for $\ave{T_{\mu \nu}(x,t)}$ where one interprets $G_{\alpha \beta \gamma \rho}$ as a (non-equilibrium finite temperature) Green's function for that operator.  In fact if one takes the Fourier transform of the integrands and realizes that the hydrostatic background implies a $\delta(k)\delta(\omega)$ in the distribution of $\ave{\tilde{T}_{\mu \nu}(k,w)}$ the usual Kubo formulae should be recovered (the perturbation $\delta \tilde{T}_{\mu \nu}$ would be a sound wave damping) provided $\lim_{k,w\rightarrow 0} \tilde{G}_{\alpha \beta \gamma \rho}=0$, which is a reasonable constraint satisfied by any local theory.

However, this will {\em not yet} be a generally covariant setup, since we agreed that we only have information of $G^{\mu \nu \alpha \beta}$ in one space-like sheet, and refoliations will generally exchange space and time.   
For example, \cite{jeon2} assumes a hydrostatic background.  A gradient expansion with an assumed criterion for $u_\mu$ (MIS, BDNK, Landau, Eckart, etc.) makes exactly such an assumption.   Expanding around various backgrounds building fluctuations could restore a remnant of covariance \cite{mullins1}, with a $G_{\alpha \beta \mu nu}$ chosen by hand to be `` foliation-independent'' (for BDNK-type theories, this involves nonlocal choices, as shown in \cite{mullins2}).
However, full covariance will be achieved if we have a foliation-independent application of the fluctuation-dissipation theorem, defined via the Kramers-Kronig relations.  This can be achieved if the volume-preserving diffeomorphism symmetry is maintained.  In this set-up stochastic evolution is entirely determined by a conservative dynamics, which, however, links not averages but an average and a variance.

The big general lesson is that making the effective theory generally covariant (a physical necessity, since entropy in the absence of curvature is a scalar \cite{basso,jacobson} and fluid local cells experience non-inertial transformations), it is {\em essential } that the propagator is also dynamical and evolved from initial conditions.   Any other approach, deriving the propagator purely from the microscopic dynamics (as using Kubo formulae in a gradient expansion), will inevitably break general covariance.    

Physically, the point here is that for extended objects, non-inertial transformations (changes in two non-inertial foliations) mean that what looks like ``two neighboring cells talking'' in one frame looks like ``a cell evolving in time'' in another.  This is perfectly compatible with direction in time being common to all causal foliations exactly because the objects are extended (another way to think about it is that ``cells'' and ``foliations'' are user defined and physics should be independent of them).  Hence, one needs a relation between space-like fluctuations and time-like evolution.  In general QFT such a relation only exists for asymptotic observables (the $i\epsilon$ prescription), but the local applicability of the fluctuation-dissipation theorem allows us to define this locally and fully via linear response theory.
Provided that the ansatz is Gaussian, this also fixes the dynamics.

All of this also profoundly affects the analysis of causality, which has been crucial in deriving various effective deterministic theories of hydrodynamics \cite{disconzi}.
Only $\ave{\left[ T_{\mu \nu}(x),T_{\mu \nu}(x')\right]}$ (propagators of observable quantities) and its counterpart for any conserved charge $J_\mu$ must be causal, something automatically enforced by the existence of generally covariant Kramers-Kronig formulae.
Objects such as $\ave{T_{\mu
\nu}(x')T_{\mu \nu}(x)}$  (space-like correlations) or $\ave{\left[ u_{\mu
}(x),u_{\nu}(x')\right]}$ (propagators of unobservable quantities) can be reinterpreted as fluctuations, and as such are permitted to be non-causal.  This is analogous to constraints in Gauge theory, which, as illustrated in \cite{kovner}, has strong non-trivial nonlocal correlations to maintain symmetries, but these correlations cannot be reduced to propagators.   The distinct scales of \eqn{scales} are then analogous to the two correlated ultraviolet divergences in Pauli-Villars regularization\cite{peskin}.  Respecting general covariance means they are not distinct scales but rather a mix, as dissipating sound modes are indistinguishable from thermal fluctuations.

There are of course ways that the current ansatz could not work and/or break down: An obvious one is that the partition function is not approximately Gaussian or stops being Gaussian in its evolution.  This is possible in the vicinity of a critical point \cite{cumulcrit}, but is otherwise unlikely, due to the central limit theorem and the relatively fast convergence to it \cite{biro}.  In principle, the Gaussian ansatz can be relaxed: Noether`s second theorem \cite{boulware} ensures that Ward identities for general diffeomorphism invariance exist for arbitrary n-particle correlators.   Combining this with ``(n-1)-th order" response theory should give a closed system of equations that would yield the equivalent of Fig. \ref{linearflux} for $\lnz$ with non-trivial Skewness, Kurtosis, and higher cumulants.  Numerically, this would be, of course, a much more involved project\footnote{Note that as we have shown general covariance can be achieved just within the Gaussian approximation.   At the moment it is not clear to us if this is true for higher cumulants or, analogously with the Schwinger-Dyson equation \cite{peskin}, higher cumulants will mix in a way that truncation breaks the general covariance.}. \textcolor{black}{For these reasons, while existing literature suggests hydrodynamic fluctuations will have a profound impact on dynamical universality classes of fluids with a phase transition \cite{skokov} (which might not be captured by effectives theories lacking background independence \cite{xinan,mullins1,mullins2,glorioso,eightfold,akash,singh}, given the connection between symmetry and universality) we can not at the moment say anything concrete in this regard. } 

 \textcolor{black}{The role of flow and the equation of state, although superficially different, is actually compatible with that of the usual hydrodynamics.  The point here is that at each point in the fluid, the ``local'' number of microstates (a scalar under general covariance) is maximized up to a fluctuation-dissipation relation (governed by $C_{\alpha \beta \gamma \rho}$).   In a Bayesian sense \cite{jaynes} (on which Gibbsian statistical mechanics is based on \cite{jaynesgibbs}), we use measurements of $T_{\mu \nu}$ within the fluid to estimate this number of microstates via a Gaussian ansatz.   Thus, for non-turbulent irrotational flow, one could choose
$d\Sigma_\mu \propto \beta_\mu$ in such a way to minimize $\mathrm{Im} G_{\alpha \beta \mu \nu}$, thus ensuring the cells co-moving with $d\Sigma_\mu$ are as stationary as possible.  Then one can use this $\beta_\mu$ to perform a Lorentz transformation and ascertain how close $\ave{T_{\mu \nu}}$ is to equilibrium and what is the equilibrium temperature.   Once one recovers flow and entropy from \eqn{lageos}, \eqn{uideal} and \eqn{equil}, using \eqn{gibbsduhem}, one can obtain the usual decomposition of \eqn{usualhydro}.  We think the result will be similar to the so-called density frame \cite{singh} and their evolution is as close as possible to reversible \cite{spalinski}.}

But since this estimation is bounded by an uncertainty given by $C_{\alpha\beta\gamma\rho}$, there is no unique way to do it, so there is an ambiguity of the temperature and $\beta_\mu,\mathcal{T},\Pi_{\mu \nu}$, but the dynamics should be invariant w.r.t. this ambiguity to the leading order, since $\ave{T_{\alpha \beta}}$ and $C_{\alpha \beta \gamma \rho}$ do not change.   What changes is the microscopic entropy, of which however we only have a Bayesian understanding.   If we ``guess the entropy right'', Eq. \ref{gibbsduhem} will give exactly the equation of state one gets from microscopic statistical mechanics \cite{huang}.  Otherwise, there will be corrections which one can think of as interactions of sound waves generated by thermodynamic fluctuations.   However, these corrections will not change the dynamics to Gaussian order in cumulants.

Furthermore, the propagator is not directly a function of the Lagrangian, and a one-to-one decomposition of it into ``transport coefficients'' and the ``equation of state'' is in principle problematic beyond the asymptotic expressions of \eqn{transcoeff} and \eqn{covarianthuang}.  For the same reason, spectral analysis of small perturbations (such as \cite{grozanal}) might be unreliable since they are not defined on every element in the ensemble but only on the ensemble as a whole.  Nevertheless, if one uses random polynomials to analyze propagation of ensembles of perturbations \cite{randompol}, it seems that indeed non-perturbative fluctuations help to propagate small perturbations beyond the Knudsen number average.

In the limit where $G_{\mu \nu \alpha \beta}$ is highly peaked, close to the $\delta-$ function limit, that is, dynamics is nearly deterministic, \eqn{ward} is equivalent to a conservation law.
In this limit, as discussed in \cite{crooks,crooks2}, an entropy conservation law emerges, leading to ideal hydrodynamics \eqn{usualhydro}  and \eqn{equil}. Thus, we see that the limit of the $\delta-$ function is smooth: while the constraint between variance and average becomes singular in \eqn{ward} as the variance disappears, the volume-preserving diffeomorphisms become a symmetry of the dynamics of the average motion, generating a conserved entropy current as well as a nonlocal conserved circulation \cite{cov3}.   The apparent divergence in \eqn{ward} can therefore be regulated by sending the number of degrees of freedom ($T_0$ in \eqn{uideal}) to infinity, or equivalently the characteristic volume scale inherent in the volume-preserving diffeomorphisms to zero.  The volume-preserving diffeomorphism symmetry is ``broken'' (as argued in works such as \cite{montenegro,grozdanov}) in the same way that any symmetry of a distribution becomes invisible when that distribution collapses to a $\delta$-function.
This is consistent with the fact that the ideal hydrodynamic limit only makes sense when fluctuations disappear and vice versa, fluctuations require some dissipation (though very different from the naive fluctuation-dissipation relation).  The analogous situation in quantum field theory is that truncating a Schwinger-Dyson equation \cite{peskin} generally gives a gauge-dependent result, but in the classical high occupation number limit this gauge dependence disappears.
All this is also consistent with the idea that the coarse graining parameter $T_0$ is implemented through the radius of curvature of the foliation.  General covariance applies for foliations smoother than this critical radius of curvature because over such scales the dynamics is purely determined by the number of microstates.   Hence, implementing general covariance w.r.t. foliations is a way of making sure the system is locally thermalized in that, over that scale, dissipative evolution is indistinguishable from fluctuations.

In the $\delta-$ function limit, the weak energy condition \cite{visser} should be directly related to the second law of thermodynamics via the Gibbs stability criteria.   The average of the Gaussian will then be related to the information current defined in \cite{gavinfo}, through the covariantized equation of state \eqn{covarianthuang} and \eqn{gibbsduhem}.  
However, unlike in \cite{gavinfo}, we do not necessarily impose the Gibbs stability at scales relevant for hydrodynamics.   In fact, when microscopic statistical mechanics is considered, dynamics is expected to be absolutely chaotic near the first scale of \eqn{scales} (where Poincaré cycles occur) and partially chaotic in the middle of the chain of inequalities in \eqn{scales} (where microscopic fluctuations and turbulent dynamics mix).
 In the usual thinking, instability are indications that the expansion is incorrect, that the theory is ill-posed.  However, a ``local instability'' in entropy (a convex $G_{\mu \nu \alpha \beta}$) could just indicate strong fluctuations that take the system out of that local equilibrium region.  The non-relativistic phenomenon of spontaneous stochasticity \cite{spont1,spont2}, which we mention toward the conclusion, could be related to this, as could the emerging Gauge symmetries recently discovered in many body systems \cite{gaugemech}.

These considerations motivate a profound revision of the criterion of applicability of hydrodynamics.    Usually, this criterion is written in terms of the ``Knudsen number'' Kn, defined as the ratio of microscopic space-time scales (viscosity $\eta$ or relaxation time $\tau_\Pi$) to gradients of microscopic scales ($\min_\beta X$ means the value of $\beta$ that minimizes $X$)
\begin{equation}
\label{knudsendef}
\text{Kn} \equiv \min_\beta \frac{\left( T^{\mu \nu}-T_0^{\mu \nu}(\beta^\mu)\right)\partial_\mu \beta_\mu}{T_\alpha^\alpha} \sim \frac{\eta}{(e+p) R} \sim \frac{\tau_\Pi}{R} \ll 1 \eqcomma \text{Kn} \geq \frac{1}{\sqrt{\text{N}_{\text{DoF}}}},
\end{equation}
where $R$ is the characteristic size of the system and $T_0^{\mu \nu}$ is given by \eqn{equil}.  Flow ambiguity \cite{disconzi,kovlec} is included in this definition, at most by the possibility of ``  choosing'' the vector $\beta$ that minimizes $\text{Kn}$ and fluctuations by some fluctuation dissipation relation that relates $\text{Kn}$ to the number of degrees of freedom $N_{\text{DoF}}$.  This criterion is exactly what is challenged by the observation of close to the ideal hydro behavior in small systems \cite{nagle}.  

The discussion here motivates a different approach: Crooks type reasoning and a Gaussian type partition function means that you are close to ideal hydrodynamics if the system is $1-\sigma$ away from equilibrium (Crooks type dynamics is, unless entropy has multiple extrema, just expected to make the system oscillate around global equilibrium in its evolution).   Looking at \eqn{gengausssolved}, this criterion is not parameterized by the Knudsen number, but by the new variable $\xi$
\begin{equation}
\label{knudsenfluct}
\xi \equiv \frac{
\min_\beta \left[ \left( \ave{T_{\mu \nu}}-T_0^{\mu \nu}(\beta^\mu)\right)\partial^\mu \beta^\mu \right]} {\sqrt{C^{'\alpha \beta}_{\alpha \beta}}} \leq 1,
\end{equation}
where $\beta$ is minimized over an ensemble of $T_{\mu \nu}$ characterized by an average and a width, $T_0$ is the equilibrium part (determined by matching ensemble averages).   In practice $\xi$ measures the difference between the \textcolor{black}{viscous} equilibrium deviation and one fluctuation within the ensemble of configurations.
\textcolor{black}{the closest counterpart to this quantity in Israel-Stewart hydrodynamics is the causality condition $\tau_\Pi \geq \eta/(e+p)$, linking the relaxation time to the ratio of viscosity to enthalpy.  Just like for perturbations on a scale higher than $c_s/\tau_\Pi$ a viscous response does not turn on, perturbations faster than the correlation length set by $\sqrt{C'}\times \ave{T_{\mu \nu}} \times \partial u$ are inherently stochastic}.

It is clear that qualitatively \eqn{knudsenfluct} depends in a completely different way from \eqn{knudsendef} on the number of degrees of freedom, since one compares the {\em deviation from equilibrium} with {\em the width}.  As the number of degrees of freedom goes to infinity, the Gibbs-Duhem relation and \eqn{covarianthuang} mean $\xi \sim Kn$, but away from this limit $\text{Kn}$ are bound to increase with the number of DoFs while $\xi$ could well decrease.   This makes it plausible that the hydrodynamics developed here can be applied where traditional hydrodynamics fails.   This means that the behavior of the Navier-Stokes corrections argued by \cite{geroch,pretorius} should be dramatically different when fluctuations are considered. 
In addition, \eqn{knudsenfluct} is also independent of both the Grad limit and the `t Hooft limit \textcolor{black}{(in fact $\xi$ diverges in both, and the general covariance and
volume-preserving diffeomorphisms do not apply). Thus, while we cannot estimate $\xi$ due to our ignorance of its numerator (it only has been calculated in the lower left panel of Fig. \ref{figscales}, when $\xi$ diverges), we can be reasonably certain that it does not depend on $N_{dof}$ in the same way as \eqn{knudsendef}, which is encouraging in view of phenomenology \cite{nagle} .}

\textcolor{black}{It is also a legitimate question whether the small parameter of this cumulant expansion is of the order of \eqn{knudsenfluct}, which would mean that beyond-Gaussian corrections are necessary to describe small systems.  While numerical simulation is necessary to answer this question with certainty, we note that the break-down of the Gaussian approximation happens when Taylor-expanding \eqn{gengauss}.   It is not an expansion in the ``gradient of the momentum fluctuation'' but rather of its expectation value in an ensemble sum.  The point here is that the Gaussian is not just a cumulant truncation but it is a very special function:
It is the only finite width distribution which is a fixed point of convolution equations \cite{renoprob} (which is the reason why Gaussian theories usually emerge as continuum limits in lattice regularization \cite{peskin}).   Given that it is a basic fact of statistical mechanics that outside of a phase transition widths scale finitely with the volume, and that the small viscosity small fluctuation limit has a fixed point in ideal hydrodynamics, it should follow that deviations from Gaussian theory of the integration measure would show up as anomalous dimension logarithmic corrections \cite{peskin} rather than as higher cumulants.  Thus, we believe that further terms in the Taylor expansion in \eqn{gengauss} does not give corrections of order $\xi^{n\geq 2}$ but rather redefines $\xi$ by corrections logarithmic in fluctuations of Lorentz generators around the co-moving frame. }  To prove this one would have to set up a renormalization group equation \cite{peskin} centered around $T_0$ in \eqn{uideal} and \eqn{lageos} as well as the Gaussian solution \eqn{gengausssolved}
\begin{equation}
T_0 \rightarrow T_0+\Delta T_0 \eqcomma s\rightarrow s+\Delta s \eqcomma F(s)\rightarrow F(s)+\Delta F(s)
\end{equation}
\[\  \left( \sqrt{ \prod_i \lambda_i \mu_i}\sqrt{ C_{\alpha \beta}^{'\alpha \beta} }\right)^{-1} = 
 \left( \sqrt{ \prod_i (\lambda_i+\Delta\lambda_i) (\mu_i+\Delta \mu_i)}\sqrt{ C_{\alpha \beta}^{'\alpha \beta}+ \Delta C_{\alpha \beta}^{'\alpha \beta} }\right)^{-1} 
\]
and study it's structure and fixed points.   This is left to a subsequent work.

Let us now make a comparison between our approach and the ``general frame global equilibrium'' proposed in works such as \cite{palermo}.
The global equilibrium is defined there as $\mathrm{Max}\left[ \ave{\ln \hat{\rho}} \right]_{\beta_\mu,\mu,...}$, but the grand canonical ensemble underlying the Zubarev statistical operator also presupposes the hydrodynamic limit, which is ill defined if $O^{-1} \nabla O \simeq 1/R,1/T$ since, in the regime of hydrodynamic turbulence, statistical fluctuations talk to macroscopic excitations, and this global equilibrium becomes unstable. However, the local equilibrium is well defined and forms a solid basis of an
EFT such as the one proposed here.
The ambiguity reflects the tension inherent in defining ``the thermodynamic limit'' (volume goes to infinity, static for a long time) in a system with local fluctuations in a way that respects relativistic causality and is due to the entropy in Global equilibrium being Boltzmannian (only ``micro'' Dofs) and not 
Gibbsian (covariantly ``coarse-grained'' Dofs, including fluctuation-generated sound waves and vortices).    The equivalent of the entropy density can then reasonably be expected to be $\ave{\lnz}$ with the average taken over all field configurations, proportional to \eqn{gengausssolved}. 

We also note that the local applicability of the Crooks theorem means that the equilibrium state is not defined by the separation of ``slow'' and ``fast'' degrees of freedom (common to gradient expansions and Schwinger-Keldysh approaches), but rather by the limit where the dynamics is dominated by microstate counting.   The system can respond to ``fast'' changes, but the response is determined by the microstates compatible with those changes.  

\textcolor{black}{The Crooks-type picture and ergodicity provide a way to justify the rather strong assumption of the existence of a generally covariant linear response, merely assumed at the start of \secref{seclin}.
First of all, as already pointed out in \cite{ergodic}, the fluid dynamics is ``special'' because local ergodicity makes the local equilibrium dynamics described here ``rigid''.  If one breaks the symmetries of hydrodynamics (a ``jelly'', a solid or other extensions explored in works such as \cite{framid,universality}), the microscopic phase space becomes locally decomposable \cite{khinchin}.   This means that while global equilibrium is well defined, local equilibrium is not because fluctuations introduce long-range correlations (``solids are brittle'', especially relativistic ones).  The symmetries of ideal fluid dynamics are what guarantee indecomposability at the level of each fluid cell. 
Generically, the linear response approach of \cite{kadanoff,tong,forster} assumes neither such symmetries nor meaningful local equilibrium, nor a generally covariant locally decomposable entropy definition stable against fluctuations.  Such calculations also neglect backreaction from microscopic to macroscopic dynamics (``micro'' and ``macro'' scales are separable).
However,  fluids, including fluids with comparably few degrees of freedom seem to exist, and the eigenstate thermalization hypothesis makes it reasonable that a Gibbsian definition of entropy localizeable on the $l_{\text{micro}}$ scale is possible.  It is then unavoidable that locally w.r.t. this scale this entropy is a scalar under general coordinate transformations. \secref{seclin} is then an effective theory procedure for imposing symmetries and seeing what constraints follow from them on the dynamics}.

Such a picture is also natural when the underlying thermalization mechanism is eigenstate thermalization \cite{berry}.  Since the maximal entanglement density matrix is also identical to the eigenstate-thermalized density matrix \cite{popescu}, and one could also see this arising under the maximum entanglement hypothesis, by postulating that in the ideal hydrodynamic limit every subvolume of the system is in a maximally entangled state under every foliation, according to the picture discussed in \cite{kharzeev} (we note here that mixing is generally not Lorentz invariant \cite{terno} but, in analogy with \cite{ergodic}, could be covariant under every foliation).

However, we caution that the relation between eigenstate thermalization and maximal entanglement is not so straightforward. 
 On one hand, it is reasonable that a chaotic quantum system is extremely susceptible to maximal entanglement, hence eigenstate thermalization and maximal entanglement will coincide.  On the other, deviations from {\em perfect} eigenstate thermalization, examined in this work, could look very different from deviations from maximal entanglement, since in one case correlations between cells are  classical Bell-type inequalities \cite{sorella} are satisfied, in the other they are quantum and they are not (though one could imagine that each cell is maximally entangled with many cells only higher order Mermin-type correlations \cite{sorella} become non-classical).  Such issues are well beyond the current work.

We conclude with possible further extensions of this work. An immediate open problem is the extension of the ideas discussed here to the hydrodynamics with spin \cite{spinreview}. The main hope is that the approach outlined here has the potential to clarify the pseudogauge issue. As shown for example in \cite{gursoy} one can formulate a spin hydrodynamics with gradient expansion using torsion perturbations to formulate spin currents.  This hydrodynamics is pseudogauge invariant at the deterministic level but the entropy current is not, and hence fluctuations will in general be pseudogauge dependent (as is seen elsewhere \cite{florkfluct}). 
Is this reasonable?  While the pseudogauge appears as a mathematical ambiguity of conserved currents \cite{jeonspin} 
 we know that physically a pseudogauge change can be thought of in terms of Noether currents as a field redefinition combined with a non-inertial coordinate transformation \cite{brauner}.  This is exactly equivalent to the way dynamics is independent of flow redefinitions once general covariance is enforced at the fluctuation level, the main theme of the present work.  Although the linear response torsional kernel equivalent of \secref{seclin} was developed in \cite{gursoy}, to our knowledge, there is no gravitational ward identity for gravity with torsion. Once that is developed, we think our arguments here can be extended with few conceptual difficulties to hydrodynamics with spin to produce an exactly pseudo-gauge invariant dynamics, although there might be surprises (given spin and causality imply dissipation \cite{montecausal}, the Gaussian propagator should have a minimum width).   A similar discussion can be made for nonlocal effects of gauge symmetries if the mean free path is smaller than the color coherence regions \cite{ghosts}.

Another interesting question is the non-relativistic limit of what we are proposing here.   Since the Galileo group is not a subgroup of the Lorentz group, one inherently needs to make additional assumptions to obtain such a limit \cite{kaminski}.   We believe that the correct assumption is incompressibility, $c_s² \rightarrow \infty$, where one gets a ``sonic Galileolimit '', where, as Vladimir Arnold has shown \cite{arnold}, the ideal hydrodynamics follows from volume-preserving diffeomorphisms.
Since time in this limit is absolute, what happens to ergodicity is still mysterious.  But we are confident that spontaneous stochasticity \cite{spont1,spont2} together with hidden symmetries of incompressible fluids \cite{fluidsym} indicate that a theory similar to the one we proposed here emerges, provided that the statistical mechanics of incompressible matter are properly understood
\cite{incompress}.

On the other end of scales, the Crooks theorem results in \cite{basso} give us hope that once curvature is included in this formalism, entropy will lose its uniqueness, but relative entropy will transform covariantly so that a fluctuation-dissipation dynamics will remain background independent.  This could provide a realization of entropic gravity \cite{jacobson} where the general covariance holds exactly.
This could lead to a quantitative implementation of the conjectures made in \cite{megrav1,megrav2}.

In conclusion, we have proposed a theory of viscous fluctuating relativistic hydrodynamics which is explicitly based only on observable conserved currents and possesses general covariance and the symmetries of ideal hydrodynamics.  As well as addressing the fundamental problem of understanding the relationship between hydrodynamics and statistical mechanics, and of global and local equilibrium, we believe this approach will give timescales of equilibration of small systems which differ considerably from usual expectations, thereby giving a new justification to the apparent hydrodynamic behavior in small systems.  Quantitative investigation of this is left to future work.

G.M.S. is supported by the CAPES doctoral fellowship 88887.648216/2021-00.
GT thanks Bolsa de produtividade CNPQ
305731/2023-8 and FAPESP 2023/06278-2 for support.
We thank Mauricio Hippert, Masoud Shokri and Sangyong Jeon for a careful reading of the manuscript and helpful discussions and suggestions, and Akash Jain for helpful comments.

\end{document}